\pdfoutput=1
\documentclass[
preprint,
prd,superscriptaddress,preprintnumbers,nofootinbib]{revtex4}
\usepackage{graphicx}
\usepackage{epsfig}
\usepackage{lipsum}
\usepackage{mathrsfs}
\usepackage{enumitem}
\usepackage{bm}
\usepackage{latexsym,amssymb,amsmath,amsfonts,amssymb,txfonts,pxfonts,wasysym,float}
\usepackage{color}
\usepackage{subfigure}

\newcommand{\beq}[1]{\begin{equation}\label{#1}}
\newcommand{\eeq}{\end{equation}}
\newcommand{\bea}[1]{\begin{eqnarray} \label{#1}}
\newcommand{\eea}{\end{eqnarray}}
\newcommand{\ba}{\begin{array}}
\newcommand{\ea}{\end{array}}

\def\be{\begin{equation}}
\def\ee{\end{equation}}
\def\gs{\mathrel{
   \rlap{\raise 0.511ex \hbox{$>$}}{\lower 0.511ex \hbox{$\sim$}}}}
\def\ls{\mathrel{
   \rlap{\raise 0.511ex \hbox{$<$}}{\lower 0.511ex \hbox{$\sim$}}}}

\newcommand{\comment}[1]{}

\usepackage[usenames,dvipsnames]{xcolor}
\definecolor{orange}{cmyk}{0,0.5,1,0}
\definecolor{rossoCP3}{cmyk}{0,.88,.77,.40}
\definecolor{graa}{rgb}{0.8,0.8,0.8}
\definecolor{blaa}{rgb}{0.2,0.2,0.6}

\begin{document}

\title{\color{rossoCP3}{Decay of multiple dark matter particles to dark radiation in different epochs does not alleviate the Hubble tension}}

\author{Luis A. Anchordoqui}

\affiliation{Department of Physics and Astronomy,\\  Lehman College, City University of
  New York, NY 10468, USA
}

\affiliation{Department of Physics,\\
 Graduate Center, City University
  of New York,  NY 10016, USA
}

\affiliation{Department of Astrophysics,
\\ American Museum of Natural History, NY
 10024, USA
}

\author{Vernon Barger}

\affiliation{Department of Physics,\\ University of Wisconsin, Madison, WI 53706, USA}

\author{Danny Marfatia}
\affiliation{Department of Physics and Astronomy,\\ University of Hawaii, Honolulu, HI 96822, USA}
\affiliation{Kavli Institute for Theoretical Physics, University of California, Santa Barbara, CA 93106, USA}

\author{Jorge F. Soriano}
\affiliation{Department of Physics and Astronomy,\\  Lehman College, City University of
  New York, NY 10468, USA
}

\affiliation{Department of Physics,\\
 Graduate Center, City University
  of New York,  NY 10016, USA
}

\begin{abstract}
  \vskip 2mm \noindent Decaying cold dark matter (CDM) has been
  considered as a mechanism to tackle the tensions in
  the Hubble expansion rate and the clustering of matter. However,
  polarization measurements of the cosmic microwave background (CMB)
  severely constrain the fraction of dark matter decaying before
  recombination, and lensing of the CMB anisotropies by
  large-scale structure set strong constraints on dark matter decaying
  after recombination. Together, these constraints make an
  explanation of the Hubble tension in terms of decaying dark matter
  unlikely. In response to this situation, we investigate whether a dark
  matter ensemble with CDM particles decaying into free streaming dark
  radiation in different epochs can alleviate the problem. We find
  that it does not.
  \end{abstract}
\maketitle

\section{Introduction}
Shortly after high-resolution experiments heralded the field of
precision cosmology, low- and high-redshift observations gave rise to a
tension in the measurement of the present-day expansion rate of the Universe ($H_0$)
and the clustering of matter ($S_8$). Assuming the standard $\Lambda$
cold dark matter (CDM) cosmological model, the Planck Collaboration
examined anisotropies in the cosmic microwave background (CMB)
temperature and polarization fields to infer that the Universe is expanding
$67.27 \pm 0.60$~kilometers per second faster every megaparsec~\cite{Planck:2018vyg},
whereas the most influential measurements of the late Universe by the SH0ES experiment, peg the Hubble constant at
$73.2 \pm 1.3~{\rm km/s/Mpc}$~\cite{Riess:2020fzl}. For a recent compilation of
other late Universe $H_0$ measurements, see e.g.~\cite{DiValentino:2020zio}. When the late
Universe measurements are averaged in different combinations, the
$H_0$ values disagree between $4.4\sigma$ and $6.3\sigma$ with the one
reported by the Planck Collaboration~\cite{Verde:2019ivm}. The
statistical significance of the mismatch between the high $S_8$ value estimated
by the Planck Collaboration assuming $\Lambda$CDM and the lower value preferred
by cosmic shear measurements is somewhat smaller  at $\sim
3\sigma$~\cite{DiValentino:2020vvd}. It is desirable that the $H_0$ and $S_8$ tensions
be addressed simultaneously, but currently none of the
proposed models have done so to a satisfactory
degree~\cite{DiValentino:2021izs,Schoneberg:2021qvd,Anchordoqui:2021gji}.

In CMB parlance, $\theta_{\rm LS} \equiv r_{\rm LS}/D_M(z_{\rm LS})$ is the
angular size of the sound horizon at the last scattering (LS) surface,
where $r_{\rm LS}$ is the linear size of the sound horizon (i.e., the comoving distance traveled by
a sound wave from the beginning of the universe until
recombination) and $D_M(z_{\rm LS}) = \int_0^{z_{\rm LS}} dz/H(z)$ is
the comoving angular diameter distance from a present day observer to
$z_{\rm LS}$, with $H(z)$ the
redshift-dependent expansion rate. Since $\theta_{\rm LS}$
can be precisely measured from the
locations of the acoustic peaks in the CMB temperature and
polarization anisotropy spectra, given $r_{\rm LS}$, an
estimate of $H_0$ follows from $D_M(z_{\rm LS})$.

Several models in which an unstable component of multicomponent CDM decays into dark radiation have been
proposed to relax the $H_0$ and $S_8$
tensions~\cite{Menestrina:2011mz,Hooper:2011aj,Gonzalez-Garcia:2012djt,Berezhiani:2015yta,Vattis:2019efj,Enqvist:2015ara,Abellan:2020pmw,Abellan:2021bpx}. These
models can be classified according to the particle's decay width
$\Gamma$. For models with short-lived particles, {\it viz.}
$\Gamma \gtrsim 10^6~{\rm Gyr}^{-1}$, CDM is depleted into dark
radiation at redshifts $z > z_{\rm LS}$, thereby increasing the expansion rate
while reducing the comoving linear size of the sound
horizon~\cite{Menestrina:2011mz,Hooper:2011aj,Gonzalez-Garcia:2012djt}. Since
the value of $\theta_{\rm LS}$ is a CMB observable that must be kept
fixed, a reduction of $r_{\rm LS}$ simultaneously decreases
$D_M(z_{\rm LS})$ and increases $H_0$. For models with long-lived
particles, CDM is depleted into radiation at $z < z_{\rm LS}$ and
 matter-dark energy equality is shifted to earlier times
than in $\Lambda$CDM, allowing for an increase in
$H_0$ at late times~\cite{Berezhiani:2015yta,Vattis:2019efj}. Furthermore, two-body
decays that transfer energy from CDM to dark radiation at redshift $z < z_{\rm LS}$
reduce the matter content in the late universe to
accommodate local measurements of $S_8$~\cite{Enqvist:2015ara,Abellan:2020pmw,Abellan:2021bpx}. For
$\Gamma \gtrsim H_0 \sim 0.7~{\rm Gyr}^{-1}$, most of the unstable
dark matter particles have disappeared by $z=3$ (with implications for
IceCube observations if sterile neutrinos play the role of dark
radiation~\cite{Anchordoqui:2015lqa,Anchordoqui:2021dls}), whereas for
$\Gamma \lesssim H_0$, only a fraction of the unstable dark matter
particles have had time to disappear.

A point worth noting is that the most recent CMB data
severely constrain the fraction of unstable dark matter in all of
these
models~\cite{Chudaykin:2016yfk,Poulin:2016nat,Clark:2020miy,Chudaykin:2017ptd,Nygaard:2020sow,Anchordoqui:2020djl}. On
the one hand, the fraction of short-lived particles is
strongly constrained by CMB polarization measurements~\cite{Nygaard:2020sow,Anchordoqui:2020djl}. On the other hand, the lack of dark matter at low redshifts
reduces the CMB lensing power which is at odds with data
from Planck~\cite{Chudaykin:2016yfk,Poulin:2016nat,Clark:2020miy}. The
inclusion of measurements of baryon acoustic oscillations (BAO) yields
even tighter constraints on the fraction of long-lived
particles~\cite{Chudaykin:2017ptd,Nygaard:2020sow}. All in all,
current bounds on the fraction of decaying
 particles in the hidden sector make a solution to the $H_0$ tension
in terms of decaying dark matter  unlikely. It remains to be seen, however, whether a
combination of these scenarios, with multiple dark matter particles
decaying in different epochs, can ameliorate this tension.

Dynamical dark matter (DDM) provides a framework to
model the decay of a dark matter ensemble across epochs~\cite{Dienes:2011ja}. In the DDM framework, dark matter stability
is replaced by a balancing of lifetimes against cosmological abundances
in an ensemble of individual dark matter components with different
masses, lifetimes, and abundances. This DDM ensemble collectively describes the observed dark matter abundance.
How observations of Type-Ia SNe~\cite{Scolnic:2017caz} can constrain ensembles
comprised of a large number of cold particle species that decay
primarily into dark radiation was explored in Ref.~\cite{Desai:2019pvs}. In this paper, we investigate
whether CDM particles decaying in
different epochs can alleviate the $H_0$ tension.

\section{Cosmology of dark matter ensembles}\label{sec:ddm}
Inferences from astronomical and cosmological observations are made
under the assumption that the Universe is homogeneous and isotropic,
and consequently its evolution can be characterized by a spatially flat
Friedmann-Robertson-Walker line element,
\begin{equation}
  ds^2 = -dt^2 + a^2(t) \ \left(dx^2 + dy^2 + dz^2\right),
\end{equation}
where $(t,x,y,z)$ are comoving coordinates and $a(t)$ is the expansion
scale factor of the universe.

The dynamics of the universe is governed by the Friedmann equation for the Hubble parameter $H$,
\begin{equation}
  H^2(a) = \frac{8 \pi G}{3} \sum_i \rho_i(a) \,,
\label{eq:Friedmann}
\end{equation}
where $G$ is the gravitational constant and the sum runs over  the
energy densities $\rho_i$ of the various components of the cosmic fluid:  dark energy
(DE), dark matter (DM), baryons ($b$), photons ($\gamma$), and
neutrinos ($\nu$). In terms of the present day value of
the critical density $\rho_{\rm crit,0} = 3 H_0^2/(8\pi G)$, the
Friedmann equation can be recast as
\begin{eqnarray}
  H^2(a) & = & H_0^2 \ \Bigg[ \Omega_b\,a^{-3} + \Omega_\gamma
    \,a^{-4} +
               \frac{\rho_\nu (a)}{\rho_{\rm crit,0}}
               \Omega_{\rm DE} \exp \left( 3 \; \int_{a}^{1} \frac{1+w}{a'} da' \right)  + \frac{\rho_{\rm DM} (a)}{\rho_{\rm crit,0}} \; \Bigg]\; ,
\label{H}
\end{eqnarray}
where $\Omega_i = \rho_{i,0}/\rho_{\rm crit,0}$ denote the present-day
density fractions, and the subscript $0$ indicates
quantities evaluated today, with $a_0 =1$. The
energy densities of non-relativistic matter and radiation scale as
$a^{-3}$ and $a^{-4}$ respectively. The scaling of $\Omega_{\rm DE}$ is usually described by an ``equation-of-state''
parameter $w \equiv p_{\rm DE}/\rho_{\rm DE}$, the ratio of the spatially-homogeneous
dark energy pressure $p_{\rm DE}$ to its energy density $\rho_{\rm DE}$. The
observed cosmic acceleration demands $w < - 1/3$. Herein we ascribe
the DE component to the cosmological constant
$\Lambda$, for which $w=-1$, and assume three families of massless (Standard Model) neutrinos. With this in mind, Eq.~(\ref{H}) can be simplified to
\begin{equation}
  H^2(a)  =  H_0^2 \Bigg[ \Omega_b\ a^{-3} + (\Omega_\gamma + \Omega_\nu)\, a^{-4} + \Omega_\Lambda
     +   \frac{\rho_{\rm DM} (a)}{\rho_{\rm crit,0}} \; \Bigg]\; .
\label{H2}
\end{equation}

We consider a hidden sector with multiple dark matter particles with different lifetimes.
The ensemble is made up of $N$ particle species $\chi_n$, with total decay widths $\Gamma_n \equiv 1/\tau_n$,
where $n = 1,2, \cdots, N$.  They decay via $\chi_n \to \psi \overline \psi$, where  $\psi$ is a massless dark sector
particle that behaves as dark radiation. The initial abundances $\rho_n (a_{\rm prod})$, are regulated by early universe
processes and are fixed at $a_{\rm prod} \ll a_{\rm LS}$, with $t_{\rm prod} \ll
\tau_{n}$, where $a_{\rm LS}$ is the scale factor at last scattering.
For simplicity, we assume that all particles in
the ensemble are cold, in the sense that their equation-of-state
parameter may be taken to be $w_{n} \approx 0$ for all $t > t_{\rm
  prod}$.

The evolution of the energy densities $\rho_n$ of each particle species
 and of the massless dark field $\rho_\psi$ are driven by the Boltzmann equations,
\begin{equation}
  \frac{d\rho_n}{dt} + 3H\rho_n  =  -\Gamma_n \rho_n
  \label{B1}
\end{equation}
and
\begin{equation}
  \frac{d\rho_\psi}{dt} + 4H\rho_\psi = \sum_{n=1}^N \Gamma_n \rho_n
  \,,
\label{B2}
\end{equation}
respectively. In Eqs.~(\ref{B1}) and (\ref{B2}) we have omitted the collision terms
associated with inverse decay processes of the type
$\psi\overline{\psi}\rightarrow\chi_n$, because their effect on
the $\rho_n$ and $\rho_\psi$ is negligible. Our goal is to solve these equations to obtain the evolution of the Hubble parameter (Eq.~\ref{H2}), which may then be used to determine the free parameters of the model by imposing the following constraints derived from
cosmological observations:
\begin{itemize}[noitemsep,topsep=0pt]
\item The baryonic matter and radiation densities~\cite{ParticleDataGroup:2020ssz},
\begin{itemize}[noitemsep,topsep=0pt]
\item $\left.\Omega_{b}h_0^2\right|_{\rm exp}= 0.02237(15)$,
\item $\left. \Omega_{\gamma} h_0^2\right|_{\rm exp}= 2.473 \times 10^{-5}
(T_{\gamma,0}/2.7255\,\mathrm K)^4$, where $T_{\gamma, 0}=2.7255(6)\,\mathrm K$ is the current temperature of
the CMB photons,
\end{itemize}
with $H_0 = 100~h_0~{\rm
  km/s/Mpc}$.\footnote{Note that we use $h_0$ for what is usually
  referred to as $h$ in the literature, as we consider $h$ to be the
  time dependent parameter naturally defined as $h(a)=H(a)/(100\,{\rm
    km/s/Mpc})$.}  A point worth noting 
 is that the baryon density inferred from Planck data is in  good
  agreement with the $\Omega_{b} h_0^2$ determination from
  measurements of the primordial
  deuterium abundance (D/H) in conjunction with big bang nucleosynthesis
  (BBN) theory~\cite{Cooke:2016rky,Mossa:2020gjc}. We have verified that  variations
  of  $\Omega_{b} h_0^2$ within observational and modeling uncertainties do not change
  our results. \item The neutrino
number density per flavor $\alpha$ is fixed by the temperature of the CMB photons,
\begin{equation}
  n_{\nu_\alpha,0} = \frac{3}{11} n_{\gamma,0} = \frac{6 \, \zeta(3)}{11
    \pi^2} T^3_{\gamma,0} \sim 113~{\rm cm^{-3}}\,.
\end{equation}
 The energy density depends on the
neutrino masses $m_{\nu}$. Under our assumption that
$m_\nu \ll T_{\nu,0} = (4/11)^{1/3} \, T_{\gamma,0}$,
\begin{equation}
\rho_{\nu_\alpha,0} = \frac{7 \pi^2}{120} \left(\frac{4}{11} \right)^{4/3} \,
T^4_{\gamma,0} \,.
\end{equation}
\item The extra relativistic degrees-of-freedom in the early
  universe is characterized by the number of ``equivalent'' light neutrino species,
\begin{equation}
N_{\rm eff} \equiv \frac{\rho_{\rm R} -
\rho_\gamma}{\rho_{\nu}}\,,
\label{neff}
\end{equation}
in units of the density of a single Weyl neutrino $\rho_\nu$, where
$\rho_{\rm R}$ is the total energy density in relativistic particles
and $\rho_\gamma$ is the energy density of
photons~\cite{Steigman:1977kc}. For three families of massless
(Standard Model) neutrinos,  $N_{\rm eff}^{\rm SM} =
3.046$~\cite{Mangano:2005cc}. Combining CMB and BAO data with
predictions from BBN, the Planck Collaboration
reported $N_{\rm eff} = 3.04\pm 0.22$ at the 95\%
CL~\cite{Planck:2018vyg},  which corresponds to
 $\Delta N_{\rm eff}   =  N_{\rm eff} - N_{\rm eff}^{\rm SM}
< 0.214$. In our model, $\rho_R = \rho_\nu +\rho_\psi$, so that
\begin{equation}
\Delta N_{\rm eff}=\frac{8}{7} \left(\frac{11}{4}\right)^{4/3}  \frac{\rho_\psi (a_{\rm
  LS})}{\rho_\gamma(a_{\rm LS})}.\label{eq:deltaneff}
\end{equation}
The above $95\%\,\mathrm{CL}$ bound requires our
model to satisfy
 $\rho_\psi (a_{\rm LS}) \alt  0.1  \ \rho_\gamma (a_{\rm LS})$.

\item Setting $\rho_{\rm DM} = \rho_\psi + \sum_{n=1}^N \rho_n$, the
  evolution of the Hubble parameter
  must accommodate a diverse set of measurements of $H (z)$  at $z \leq 2.35$, described in more detail in \S\ref{sec:data}. While some of those measurements are independent of any cosmological model, others rely on BAO data, and a prior
on the radius of the sound horizon evaluated at the end of the
baryon-drag epoch ($r_d$) ought to be imposed. This value may be
separately obtained from a model dependent analysis of early universe
CMB data ($r_{d,e}$), or from model independent parameterizations
constrained by low redshift probes ($r_{d,l}$). We employ the
measurement for $r_{d,e}$ in model 1 (base $\rm\Lambda CDM$
model with $N_{\rm eff}$)~\cite{Anchordoqui:2021gji}:
$r_{d,e}=(148.3\pm1.9)\,\mathrm{Mpc}$. For the local universe
measurement, we use $r_{d,l} = (137 \pm 3^{\rm stat}\pm 2^{\rm
  syst})~{\rm Mpc}$~\cite{Arendse:2019hev}.\footnote{Note that the values of
$\Omega_b h_0^2$ obtained for model 1
of~\cite{Anchordoqui:2021gji} are consistent at the $1\sigma$ level with
 the Particle Data Group value~\cite{ParticleDataGroup:2020ssz}. We also note that the limits on
 $\Delta N_{\rm eff}$ from the
analysis of model 1 of~\cite{Anchordoqui:2021gji} are more restrictive
than our adopted bound, $\Delta N_{\rm eff} < 0.214$, because BBN considerations (which relax
the bound) were not taken into account in the analysis
of~\cite{Anchordoqui:2021gji}. To be conservative, we use the
bound reported by the Planck Collaboration~\cite{Planck:2018vyg}.}
\end{itemize}

\section{Setting up the system of Boltzmann equations}
In order to study the low redshift behavior of the Hubble parameter,
we need to solve the system of first order non-linear differential
equations formed by the \mbox{$N+1$} Boltzmann equations for the dark sector,
together with the Friedmann equation. Although Eq.~(\ref{B1}) can be analytically reduced to $\rho_n
a^3 \exp(\Gamma_n t)= {\rm constant}$, this does not provide an advantage in solving the system, since $t\sim\int da/a H$ and $H$ is a function of $\rho_n$ and $\rho_\psi$. We therefore proceed to a fully numerical solution of the problem.

We ease this task by defining $\tilde\rho_i\equiv\rho_i/\rho_{{\rm crit},0}$ and $\tilde \Gamma_i\equiv\Gamma_i/({100\,{\rm km/s/Mpc}})$ to render the equations and free parameters dimensionless. Also, we use $u\equiv\ln a$ as an independent variable. This allows to rewrite Eqs.~(\ref{H2}), (\ref{B1}) and (\ref{B2})  as
\begin{subequations}
  \begin{equation}
    \frac{d \tilde\rho_n}{du}+3 \tilde\rho_n+\frac{\tilde\Gamma_n \tilde\rho_n}{ \,h(u)}=0\,,
  \end{equation}
  \begin{equation}
    \frac{d \tilde\rho_\psi}{du}+4 \tilde\rho_\psi-\frac1{ \,h(u)}\sum_{n=1}^N{\tilde\Gamma_n \tilde\rho_n}=0\,,
  \end{equation}
  \begin{equation}
    h^2(u) = h_0^2\left(\Omega_b\, e^{-3u}+\Omega_r\,e^{-4u}+\Omega_\Lambda+\tilde\rho_{\rm DM}(u)\right)\,.\label{eq:system-h}
  \end{equation}
\label{eq:system}
\end{subequations}
The system of equations must be supplemented with $N+1$ initial conditions, i.e., $u_{\rm prod}$.
We define $\tilde\rho_{1,\rm prod}\equiv \tilde\rho_1(u_{\rm prod})$,
and assume that the production of dark radiation in the very early universe is negligible, so that $\tilde\rho_\psi(u_{\rm prod})=0$.

It is worth mentioning that Eq.~(\ref{eq:system-h}) satisfies $h(0)=h_0$ only if $\Omega_b+\Omega_r+\Omega_\Lambda+\Omega_{\rm DM}=1$, where $\Omega_{\rm DM}\equiv\tilde\rho_{\rm DM}(0)$. This is not the case for arbitrary initial conditions so this consistency condition must be imposed after solving the system of equations. To do so, we first set $\Omega_b\,h_0^2$ and $\Omega_r\,h_0^2$ to their measured values, fix all model parameters save one, and then vary the remaining parameter until the consistency condition is met.  We choose the initial density
$\tilde \rho_{1,{\rm prod}}$ to be determined by the consistency condition, and is given by the root of the function,
\begin{equation}
\mathcal G(\tilde\rho_{1,{\rm prod}})\equiv
\alpha + h_0^2\left(\Omega_\Lambda+\Omega_{\rm DM}-1\right),
\end{equation}
where $\left.\alpha\equiv\Omega_b\,h_0^2+\Omega_r\,h_0^2\right|_{\rm exp}\approx0.0224$.  Note  that $\tilde\rho_{\rm DM,0}$  is implicitly dependent on $\tilde\rho_{1,{\rm prod}}$.

\section{Observational datasets and statistical methodology}

\subsection{The data}\label{sec:data}

 In the following we provide a succinct description of the data we use to constrain the dark matter ensembles.

\subsubsection{Supernovae magnitudes}\label{sec:data-sn}
We use the Pantheon Sample~\cite{Scolnic:2017caz}, consisting of a combination of high quality measurements of supernovae spectrally confirmed to be type Ia, and cross calibrated between different experiments to reduce systematics. Specifically, we use $(z,m_B)$ data from 1048 supernovae with $z\in[0.01,2.26]$ to constrain the luminosity distance,
\begin{equation}
  D_L(z)\equiv (1+z)\,c\int_0^z\frac{d z'}{H(z')}\,,
 \end{equation}
  which may also be written as
  \begin{equation}
  d(z)\equiv (1+z) \int_0^z\frac{dz'}{h(z')}\,,
\end{equation}
and are related by $d(z)=D_L(z) 100\,\mathrm{km/s/Mpc}/c$. In terms of the distance modulus,
$D_L=10^{1+(m_B-M_B)/5}\,\mathrm{pc}$, where $m_B$ and $M_B$ are the apparent and absolute magnitudes of the source. This may be rewritten as $m_B=M_B+A+5\log d(z)$,
with $A=5\log [c/( {\mathrm{m\,s^{-1}}})]$.

\subsubsection{Hubble parameter}
As a direct measurement of the Hubble parameter at low redshift, we use  Observational Hubble Data (OHD) inspired by Table III of \cite{Cai:2021weh}. The data use the relative ages of nearby (in $z$) galaxies, to obtain an approximation to $dz/dt$, from which the Hubble parameter can be estimated as described in Ref.~\cite{Jimenez:2001gg}. The measurements are solely dependent on models that describe the spectral evolution of stellar populations and, therefore, independent of any cosmological model. The data we use is derived using the model in \cite{Bruzual:2003tq}, and contains 30 data points from Refs.~\cite{Jimenez:2003iv,Simon:2004tf,Stern:2009ep,Moresco:2012jh,Zhang:2012mp,Moresco:2015cya,Moresco:2016mzx} with $z\in[0.07,1.965]$.

\subsubsection{Large scale structure}
 We include the large scale structure information in BAO data. Following \cite{Cai:2021weh}, we use data from the 6dF Galaxy Survey~\cite{Beutler:2011hx}, the Sloan Digital Sky Survey (SDSS)~\cite{Ross:2014qpa,BOSS:2016zkm,Ata:2017dya,deSainteAgathe:2019voe,Blomqvist:2019rah,Bautista:2020ahg,Gil-Marin:2020bct,deMattia:2020fkb,Tamone:2020qrl,Neveux:2020voa,Hou:2020rse,duMasdesBourboux:2020pck}, and Dark Energy Survey (DES)~\cite{DES:2021esc}, totaling 35 measurements with $z\in[0.106,2.35]$. The quantities $r_d/D_V$, $D_V/r_d$, $D_A/r_d$, $D_H/r_d$, $D_M/r_d$ and $H\,r_d$ are directly measured from BAO data. The different distances are related by $D_H=c/H$, $D_M=D_L/(1+z)$, $D_A=D_L/(1+z)^2$, and $D_V^3=zD_M^2D_H$. These data points must be supplemented by an experimental value for $r_d$ which is either the local ($r_{d,l}$) or the early universe ($r_{d,e}$) value, introduced in \S\ref{sec:ddm}.

 \subsubsection{Hubble constant}
 As a last data point, we include the expansion rate of the local
 ($z\approx0$) Universe,
 $H_0=(73.2\pm1.3)\,\mathrm{km/s/Mpc}$~\cite{Riess:2020fzl}. This
 data point is only included in the analysis with the local value
 $r_{d,l}\approx137\,\mathrm{Mpc}$.\footnote{We verified that
   employing the very
   recent estimate, $H_0 =  73.04 \pm 1.04~\mathrm{km/s/Mpc}$~\cite{Riess:2021jrx}, does not
   modify our results.}

\subsection{The likelihood}\label{sec:likelihood}
We now introduce the ingredients of our data analysis. Assuming Gaussian errors in the measurements, we write the likelihood of the data given the cosmological model as
\begin{equation}
L=\mathcal A\prod_{p=1}^4 \exp\left(-\frac{\chi_p^2}{2}\right)\,,
\end{equation}
where the partial chi-squared,
\begin{equation}\chi_p^2=\sum_{i=1}^{\mathcal N_p}\left(\frac{y_{p,i}-\mathcal F_p(z_{p,i})}{\sigma_{p,i}}\right)^2\,,
\end{equation}
is obtained from the data of the $p^{\rm th}$ combined sample, and $\mathcal A$ is a normalization constant which depends on whether the local or the early universe value of $r_d$ is used. Specifically, $\log\mathcal A_l\approx1208$ and $\log\mathcal A_e\approx 1212$. These combined samples are obtained from the original data variables (collectively called $x$ below). In the following, $k\equiv 100\,\mathrm{km/s/Mpc}$:
\renewcommand{\arraystretch}{1.5}
\begin{enumerate}
\item $(z,h)$ from OHD and BAO, with $\mathcal F_1(z)=h(z)$:
\begin{center}
  \begin{tabular}{c c c}
\hline \hline
  ~~~~~~~  $\quad \quad x\quad \quad $ ~~~~~~~ & ~~~~~~~$\quad \quad y_{1,i}\quad   \quad $~~~~~~~ &
                     ~~~~~~~
                                                                                                     $\sigma_{1,i}^2$
                                                                                                     ~~~~~~~
    \\\hline
    $H$ & $x_i/k$ & ${\sigma_x}_i^2/k^2$\\
    $H\,r_d$& $ \frac{x_i}{k\,\tilde r_d}$ & $(k\,\tilde r_d)^{-2}\left({\sigma_x}_i^2+\frac{x^2}{\tilde r_d^2}\sigma_{\tilde r_d}^2\right)$\\
    $D_H/r_d$& $ \frac{c}{k\,\tilde r_d\,x_i}$&$
                                                \left(\frac{c}{k\,\tilde
                                                r_d\,x_i}\right)^2\left(\frac{{\sigma_x}_i^2}{x_i^2}+\frac{\sigma_{\tilde
                                                r_d}^2}{\tilde
                                                r_d^2}\right)$ \\
    \hline \hline
  \end{tabular}
  \end{center}
\item $(z,m_B)$ from Pantheon, with $\mathcal F_2(z)=M_B+A+5\log d(z)$:
\begin{center}
\begin{tabular}{c c c}
  \hline \hline
~~~~~~~  $\quad \quad x\quad \quad $ ~~~~~~~ & ~~~~~~~$\quad \quad
                                               y_{2,i}\quad \quad
                                               $~~~~~~~ &
                                                          ~~~~~~~$\quad \quad \sigma_{2,i}^2\quad \quad $~~~~~~~ \\\hline
  ${m_B}$ & ${m_B}_i$ & ${\sigma_x}_i^2$\\
  \hline
  \hline

\end{tabular}
\end{center}

\item $(z,d)$ from BAO, with $\mathcal F_3(z)=d(z)$:
\begin{center}
\begin{tabular}{c c c}
\hline \hline
  $x$ &$\quad \quad \quad \quad y_{3,i}\quad \quad \quad \quad $ & $\sigma_{3,i}^2$\\\hline
  $D_M/r_d$ & $ (1+z_i)\, \frac{k\,\tilde r_d}{c}\,x_i$ & $ \left[(1+z_i)\,\frac kc\right]^2\left(x_i^2\sigma_{\tilde r_d}^2+\tilde r_d^2{\sigma_x}_i^2\right)$\\
  $D_A/r_d$& $ (1+z_i)^2\,\frac{k\,\tilde r_d}{c}\,x_i$ & $
                                                          \left[(1+z_i)^2\,\frac
                                                          kc\right]^2\left(x_i^2\sigma_{\tilde
                                                          r_d}^2+\tilde
                                                          r_d^2{\sigma_x}_i^2\right)$\\
  \hline
  \hline
\end{tabular}
\end{center}

\item $\left(z,(d^2/h)^{1/3}\right)$  from BAO, with $\mathcal F_4(z)=\left[d^2(z)/h(z)\right]^{1/3}$:
\begin{center}
\begin{tabular}{c c c}
\hline \hline
  $x$ &$\quad \quad \quad \quad y_{4,i}\quad \quad \quad \quad $ &
                                                                   $\sigma_{4,i}^2$\\\hline
  $D_V/r_d$ & $ \left(\frac{(1+z)^2}{z}\right)^{\frac13}\frac {k\,\tilde r_d\,x_i}{c}$ & $\left(\frac{(1+z)^2}{z}\right)^{\frac23}\frac {k^2}{c^2}\left(x_i^2\sigma_{\tilde r_d}^2+\tilde r_d^2{\sigma_x}_i^2\right)$\\
  $r_d/D_V$ & $ \left(\frac{(1+z)^2}{z}\right)^{\frac13}\frac {k\,\tilde r_d}{c\,x_i}$ & $\left(\frac{(1+z)^2}{z}\right)^{\frac23}\frac {k^2}{c^2x_i^2}\left(\sigma_{\tilde r_d}^2+\frac{\tilde r_d^2}{x_i^2}{\sigma_x}_i^2\right)$\\
  \hline
  \hline
\end{tabular}
\end{center}
\end{enumerate}
\renewcommand{\arraystretch}{1}

\subsection{Ambiguity in the choice of $\bm{N}$}\label{sec:ambiguity}
We consider $N$ to be a fixed parameter that specifies the model, while the other parameters can vary within each model.
However, the choice of $N$ can become somewhat ambiguous depending on the values taken by the other parameters.
For example, a model with $N=1$ is not distinguishable from one with $N=100$  in which all but one of the initial conditions are small enough to make their evolution inconsequential. Although the ontology of these models is different, they would be indistinguishable. To resolve this ambiguity, we enforce constraints on the free parameters such that, once $N$ is chosen, all the $N$ fields are of relevance, in the sense described below.

The field corresponding to $\tilde\rho_n$ appears in the system in two ways: as a term in the total energy density, and as a source in the equation for $\tilde\rho_\psi$. A field is \emph{directly relevant} at $u$ if its contribution to the energy density at $u$ is not negligible, and \emph{indirectly relevant} at $u$ if its contribution to $\rho_\psi$ at $u$ is not negligible. We assign a zero prior to points in the parameter space for which at least one field is both directly and indirectly irrelevant globally (which holds for almost all $u$), with the goal of assigning definite meaning to the specification of $N$.

We consider the $k^{\rm th}$ field to be directly irrelevant globally if it becomes directly irrelevant within the very early evolution of the system. This may be caused by either a low initial density or a high decay rate. For the first case, we define the first irrelevance condition,
\begin{equation}
\mathcal C_1^{(k)}:\tilde\rho_{k,\mathrm{prod}}/\tilde\rho_{1,{\mathrm{ prod}}}<\varepsilon_\rho\,.
\end{equation}
We take $\tilde\rho_{1,\mathrm{prod}}$ as a reference because it has the lowest decay rate, making it most relevant in the long term. Therefore, if a field's density is initially small with respect to $\tilde\rho_1$, it will always become smaller as the system evolves.

Unless the lowest decay rate $\tilde\Gamma_1$ is very high, the Universe will initially be dominated by radiation. In this regime, the Hubble parameter is $h=h_0\sqrt{\Omega_r}a^{-2}$, and the densities evolve as
\begin{subequations}
\begin{equation}
\rho_i(a)=\rho_{i,\mathrm{prod}}\left(\frac{a_\mathrm{prod}}{a}\right)^3e^{-\beta_i\left(a^2-a_\mathrm{prod}^2\right)}\,,
\end{equation}
\begin{equation}
\rho_\psi(a)=a^{-4}\int_{a_\mathrm{prod}}^aa'^3\mathcal F(a')\,da'\,,\label{eq:rhopsirad}
\end{equation}
\end{subequations}
where $\beta_i\equiv{\tilde\Gamma_i}/{2\sqrt{\Omega_r\,h_0^2}}$ and
\begin{equation}
\mathcal F(a)\equiv\frac{2 a_\mathrm{prod}^3}{a}\sum_i\beta_i\,\rho_{i,\mathrm{prod}}\,e^{-\beta_i(a^2-a_\mathrm{prod}^2)}\,.
\end{equation}
This allows to establish $\beta_k a_{\rm {prod}}^2$ as a measure of how quickly a field decays initially, and define the second irrelevance condition,
\begin{equation}
\mathcal C_2^{(k)}:\beta_k a_{\rm {prod}}^2>\varepsilon_\beta\,.
\end{equation}
If either of $\mathcal C_1^{(k)}$ or $\mathcal C_2^{(k)}$ holds,
$\tilde\rho_k$ is considered directly irrelevant globally.

The definition of indirect irrelevance requires more nuance, since a field may be directly irrelevant (by having a small initial energy density or a very large decay rate) and still modify the evolution of $\rho_\psi$ significantly.

\begin{figure}\centering
\subfigure{\includegraphics[height=4.6cm]{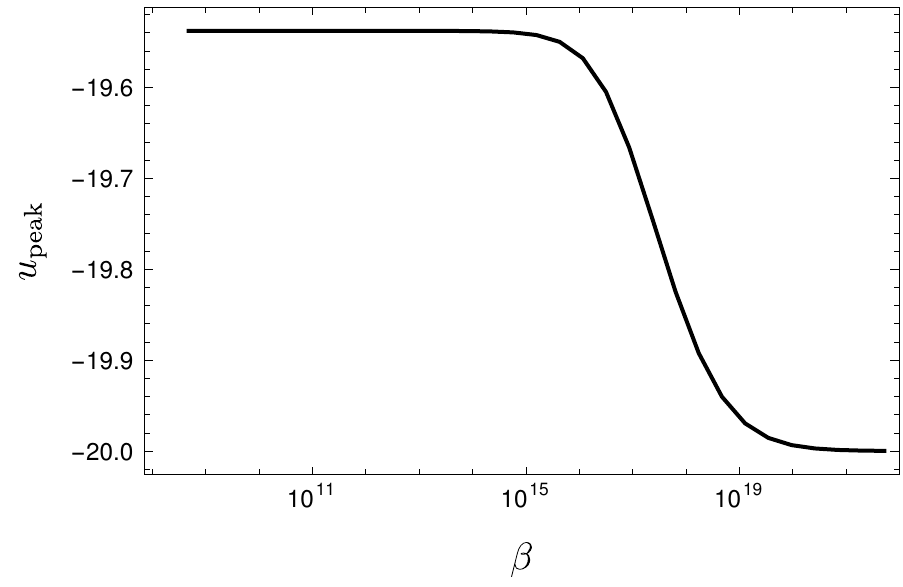}\label{fig:gpeaks}}
\subfigure{\includegraphics[height=4.6cm]{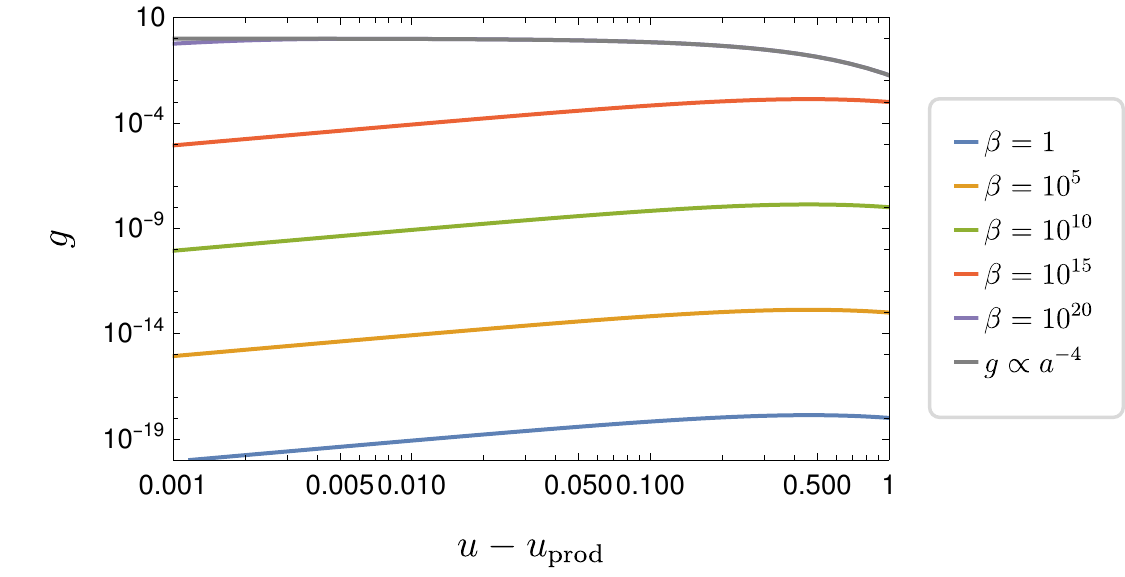}\label{fig:gcurves}}
\caption{Peak locations of $g_i(a)$ for $u_\mathrm{prod}=-20$ (\emph{left}), and the evolution of the contributions to $\tilde\rho_\psi$ by different  $\tilde\rho_i$ for several values of $\beta_i$ (\emph{right}).}
\label{fig:g}
\end{figure}

After integration, Eq.~(\ref{eq:rhopsirad}) can be written explicitly in terms of the contributions of each $\tilde\rho_i$ to $\tilde\rho_\psi$ as $\tilde\rho_\psi=\sum_ig_i(a)$, with
\begin{equation}
g_i(a)\equiv\tilde\rho_{i,p}\,\left(\frac{a_{\rm {prod}}}{a}\right)^4\left[
1-\frac{a}{a_{\rm {prod}}}e^{-\beta_i(a^2-a_{\rm {prod}}^2)}+
\sqrt{\frac{\pi}{\beta_i}} e^{\beta_ia_{\rm {prod}}^2} \frac{\mathrm{erf}\left(\sqrt{\beta_i}a\right)-\mathrm{erf}\left(\sqrt{\beta_i}a_{\rm {prod}}\right)}{2a_{\rm {prod}}}
\right]\,.
\end{equation}
These functions peak at some $a\in[a_{\rm {prod}},2^{2/3}a_{\rm {prod}}]$, depending  on the value of $\beta_i$. As can be seen from the left panel of Fig.~\ref{fig:g}, the peaks approach $a_{\rm {prod}}$ for large $\beta_i$, and $2^{2/3}a_{\rm {prod}}$ for small $\beta_i$. This reflects the fact that fields with large decay rates quickly transfer all their energy to $\rho_\psi$, which then decays as $a^{-4}$. For fields with low decay rates, the continuous energy injection to $\rho_\psi$ makes the decay slower than $a^{-4}$. A comparison of these contributions for different combinations $(\tilde\rho_{i,p},\beta_i)$ is shown in the right panel of Fig.~\ref{fig:g}.

After all contributions pass their maxima at $a=2^{2/3}a_{\rm {prod}}$, the ratio $g_i(a)/g_1(a)$ of the contributions from $\tilde\rho_i$ and $\tilde\rho_1$ to $\rho_\psi$ always decreases, since $\tilde\rho_1$ corresponds to the slowest decaying field. We can therefore use the ratios $g_k/g_1$ to discriminate relevant from irrelevant fields, in the indirect sense. We say that a field is indirectly irrelevant if
\begin{equation}
\mathcal C_3^{(k)}:\left.\frac{g_k(a)}{g_1(a)}\right|_{a=2^{2/3}a_{\rm {prod}}}<\varepsilon_g\,,
\end{equation}
which allows discrimination between models in which some field contributions to $\rho_\psi$ become negligible very early.

Besides the conditions mentioned above, there is one additional way in which the value of $N$ is ambiguous. If there are multiple fields for which $\beta_i\,a_{\rm {prod}}^2>1/\varepsilon_\beta$ and they reach their maximum contribution to $\rho_\psi$ very early, their overall contribution to $\tilde\rho_\psi$ is
\begin{equation}
\sum_i g_i(a)=\left(\frac{a_{\rm {prod}}}{a}\right)^4\sum_i\tilde\rho_{i,p}\,,
\end{equation}
which is indistinguishable from a single field with initial density $\sum_i\tilde\rho_{i,p}$ and a very large decay rate.
We therefore impose an additional condition, which is that at most one field has a decay rate such that $\beta_i\,a_{\rm {prod}}^2>1/\varepsilon_\beta$. Since the largest decay rate is that of $i=N$, this condition can be expressed as
\begin{equation}
\mathcal C_0:\beta_{N-1}a_{\rm {prod}}^2< \varepsilon_\beta\,.
\end{equation}

These conditions can be implemented as a prior in the Bayesian method described below, so that, for each $N$, we exclude the regions of the parameter space that contain at least one field that is both directly and indirectly irrelevant. Said differently, if either $\mathcal C_0$ is true, or if there exists a $k$ such that
$(\mathcal C_1^{(k)} \lor \mathcal C_2^{(k)}) \land \mathcal C_3^{(k)}$ is
true, then we assign a null prior to the model. For the models studied below, we choose conservative conditions with $(\varepsilon_\rho,\varepsilon_\beta,\varepsilon_g)=(10^{-25},2\times10^{5},10^{-10})$.

\subsection{The Markov Chain Monte Carlo}
We use a Markov Chain Monte Carlo (MCMC) Bayesian method to study how the data constrains the model parameters. We implement an adaptive Metropolis algorithm as described in Ref.~\cite{haario:2001}, in which a fixed proposal distribution is used for some small number of steps at the beginning of the chain, after which the covariance matrix of all previously sampled points is used as the covariance matrix for a multivariate normal proposal distribution. This allows the chain to adapt to the vastly different variances along different dimensions.

We consider a prior comprised of bounded uniform distributions on the cosmological parameters $h_0$ and $\Omega_\Lambda$, and $M_B$, in the intervals in Table~\ref{tab:priors}. The lower limit on $h_0$ and the upper limit on $\Omega_\Lambda$ are to ensure positivity of the energy densities.
\begin{table}[t]\centering
\caption{Prior ranges for $h_0$, $\Omega_\Lambda$ and $M_B$. \label{tab:priors}}
  \begin{tabular}{c c}
    \hline
    \hline
~~~~~~~~~~~~~~~~~~~~ parameter ~~~~~~~~~~~~~~~~~~~~ &
                                                      ~~~~~~~~~~~~~~~~~~~~ range ~~~~~~~~~~~~~~~~~~~~ \\\hline
    $h_0$   &  $[\sqrt\alpha,1]$\\
    $\Omega_\Lambda$    & $[0,1-\alpha/h_0^2]$\\
    $M_B$ & $[-25,-15]$ \\
    \hline
    \hline
  \end{tabular}
\end{table}

The priors on the initial conditions and the decay rates are also uniform with bounds  chosen to implement the relevance conditions defined in \S\ref{sec:ambiguity}.

\section{Numerical Analysis}

\begin{figure}\centering
  \includegraphics[width=0.7\linewidth]{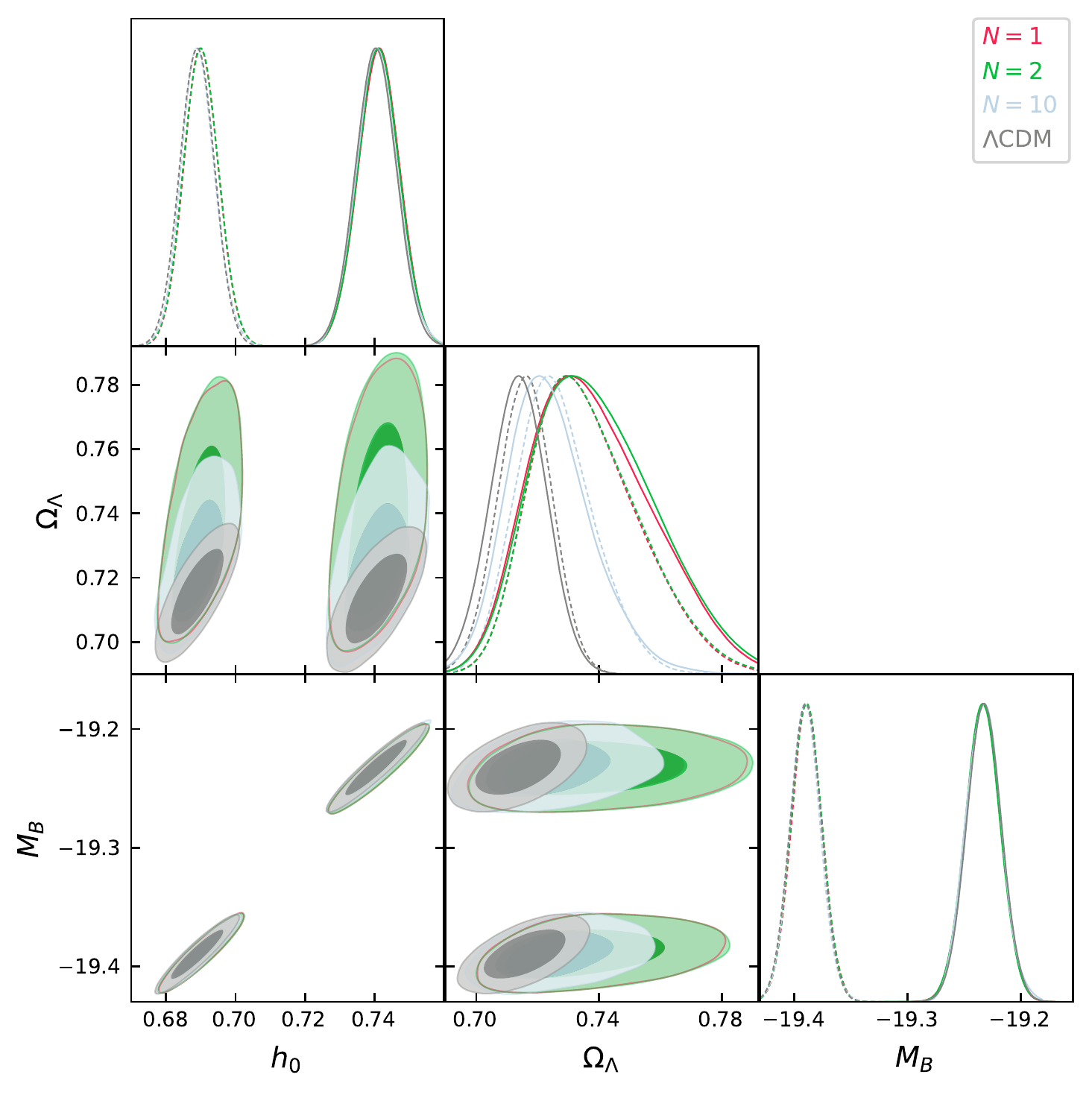}\caption{1D and 2D 68\% CL and 95\% CL posterior distributions for $h_0$, $\Omega_\Lambda$ and $M_B$ for the $\Lambda$CDM and $N=1,2,10$ models. In the 1D posteriors, the solid lines are for $r_{d,l}$ and the dashed ones for $r_{d,e}$.  These two values produce separate islands in the 2D contour plots, with lower $h_0$ and $M_B$ for $r_{d,e}$ than for $r_{d,l}$. This clearly shows how the discrepancies in $H_0$, $M_B$ and $r_d$ are related.}\label{fig:tri1} \end{figure}

We begin by presenting the results of the fit for a few cases of relatively low $N$ (1, 2 and 10) with
$\Lambda$CDM as the baseline model. We first show the posterior distributions for the cosmological parameters $h_0$ and $\Omega_\Lambda$, together with $M_B$, which are common to all models. The 1D and 2D 68\% CL and 95\% CL distributions are shown in Fig.~\ref{fig:tri1}. Clearly, none of the models differ from
$\Lambda$CDM in their predictions for $h_0$ and $M_B$, while predictions for $\Omega_\Lambda$ differ significantly. A large increase in $\Omega_\Lambda$ occurs for $N=1$, a case in which the only decaying field has to account for all the dark matter during the evolution. Nonzero decay rates produce low values of $\Omega_{\mathrm DM}$ unless the initial density for the field is high, which is incompatible with the early universe data. Therefore, there is more room for dark energy. As the number of fields increases, the decay rate of the slowest decaying field decreases, allowing for an overall increase in $\Omega_{\rm DM}$.

\begin{figure}\centering
\includegraphics[width=0.55\linewidth]{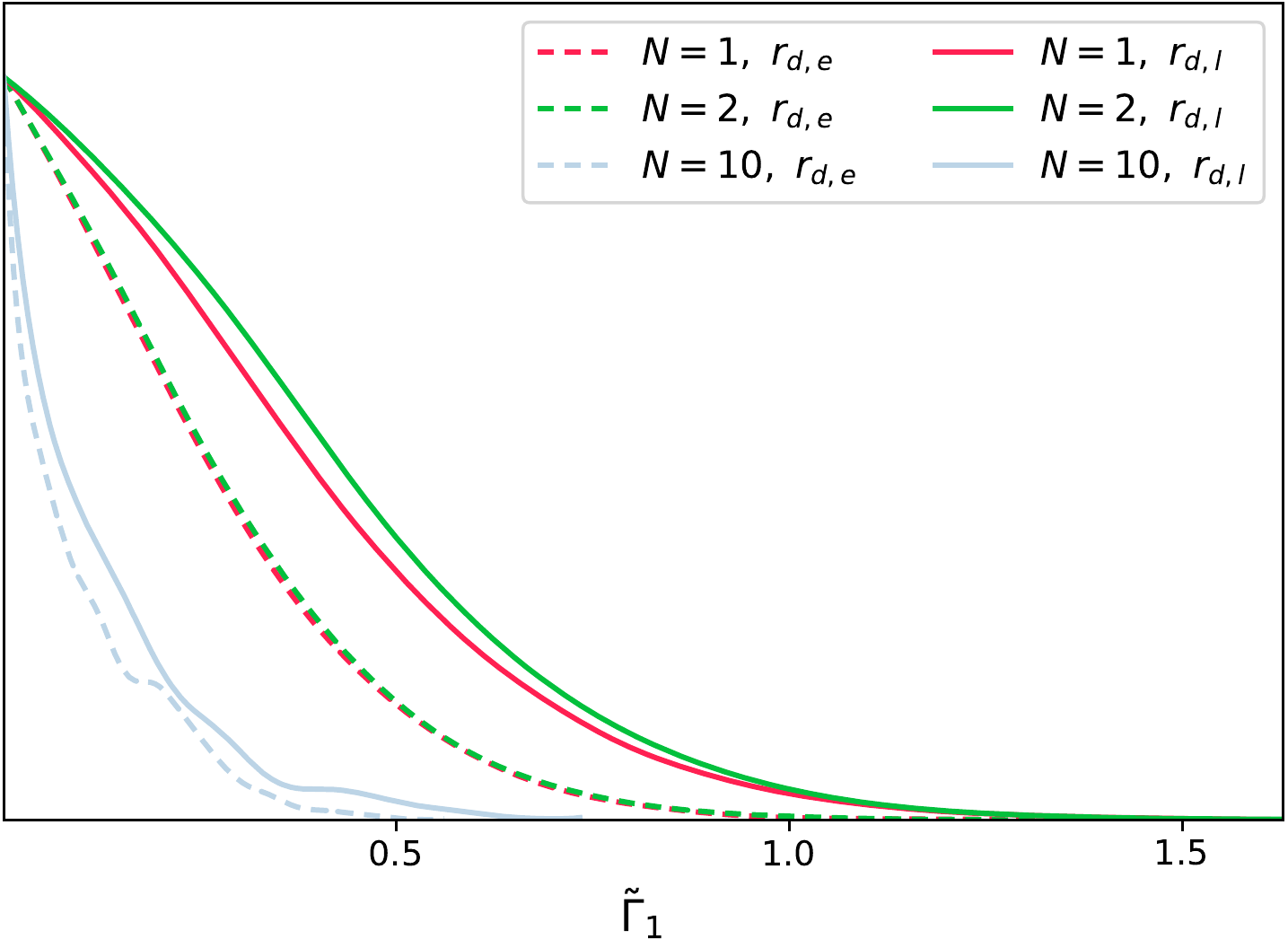}\caption{Decay rates of the slowest decaying field for $N=1,2,10$ and both values of $r_d$.}\label{fig:gammas}
\end{figure}

Regarding the model parameters (decay rates and initial densities), the results point to a slowly decaying field and a collection of fields decaying in the very early universe. In Fig.~\ref{fig:gammas} we show the decay rate $\tilde\Gamma_1$ of the slowest decaying field. Note that by definition, a value of order unity is approximately the age of the Universe.

In the $N=2$ case, there is an additional decaying field besides the slowly decaying field of Fig.~\ref{fig:gammas}. The posterior distribution for its decay rate becomes flat at its maximum value, so that the field decays in the very early Universe. Thus, we conclude that with two fields, one has a decay rate close to zero, while the other has a vanishing lifetime. This explains why the results for
$N=1$ and $N=2$ in Fig.~\ref{fig:tri1} are so similar.

The posterior distributions for the parameters in the $N=10$ case show more structure, due to the imposition of the relevance conditions on the decay rates. Nevertheless, we decide not to include them here since Fig.~\ref{fig:tri1} already shows that the model cannot affect $h_0$ significantly.

\begin{figure}\centering
\includegraphics[width=1\linewidth]{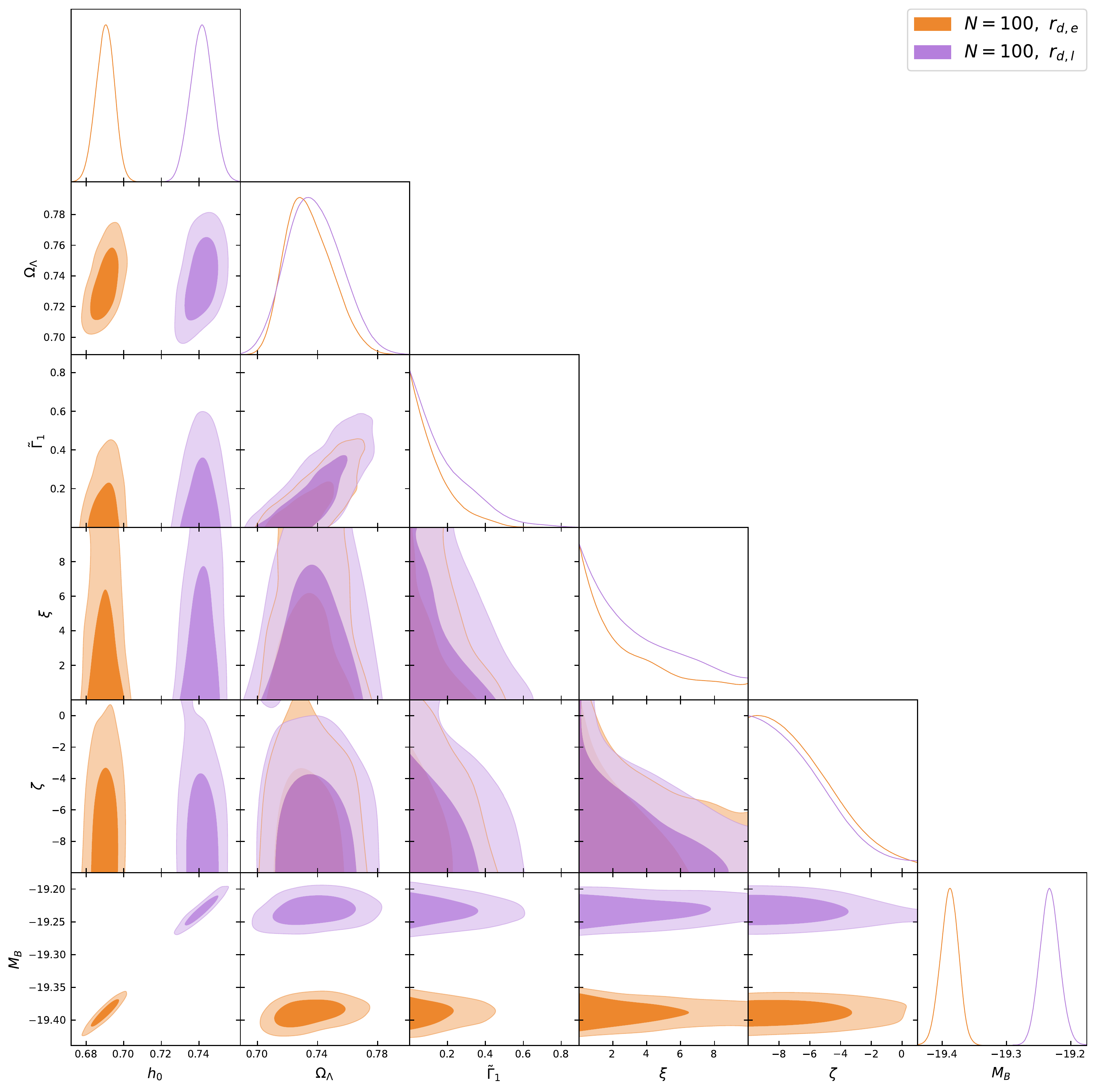}\caption{1D and 2D posterior distributions for all free parameters of the $N=100$ model.}\label{fig:tri-n100}
\end{figure}

\begin{figure}\centering
\includegraphics[width=0.7\linewidth]{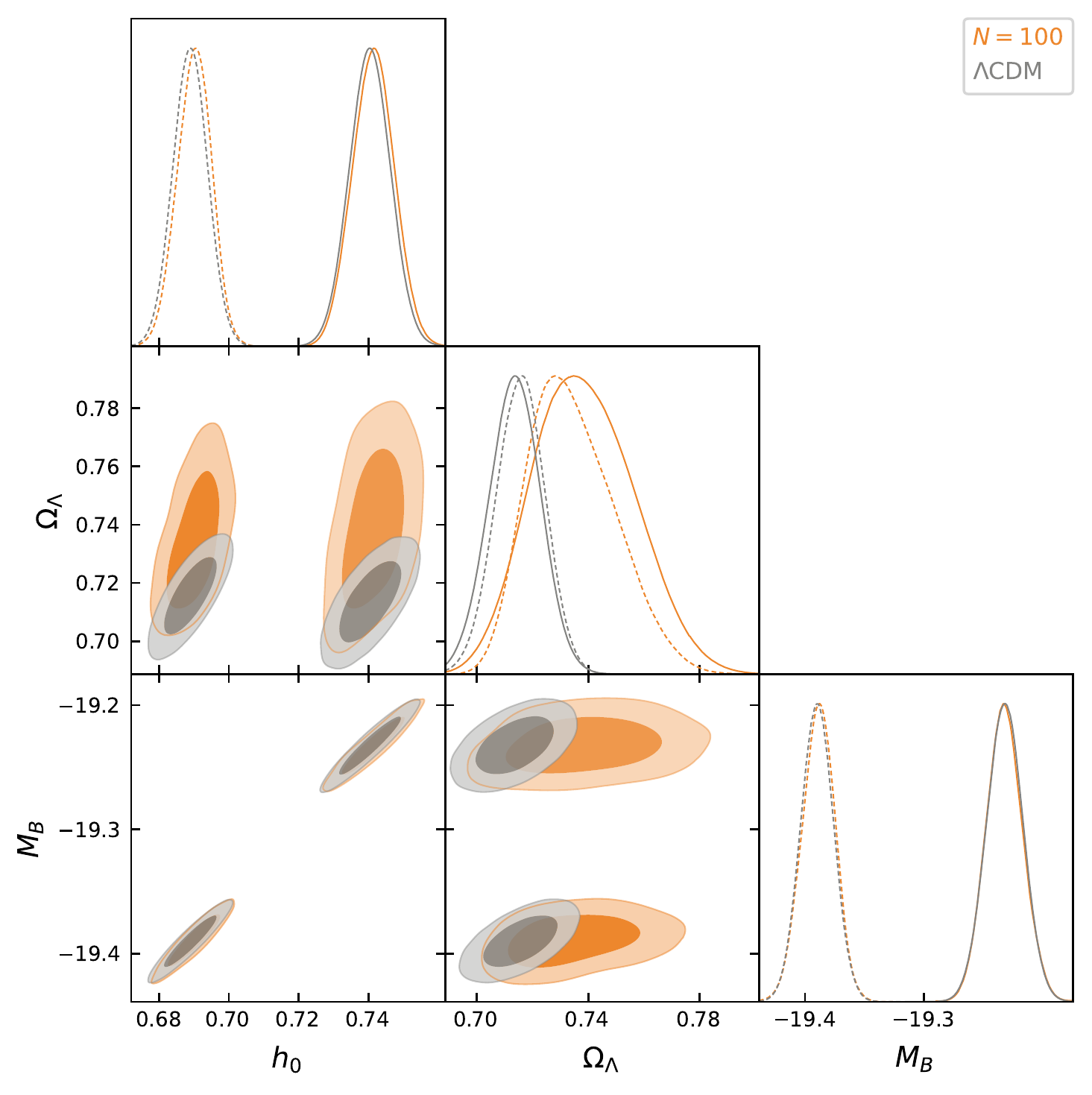}\caption{1D and 2D posterior distributions for $h_0$, $\Omega_\Lambda$ and $M_B$ for $\Lambda$CDM and $N=100$. In the 1D posteriors, the solid lines are for $r_{d,l}$ and the dashed ones for $r_{d,e}$.  
}\label{fig:cdmvsn100}
\end{figure}

We now consider a large ensemble of fields and a simple parametrization for their initial densities and decay rates~\cite{Desai:2019pvs}:
\begin{subequations}
  \begin{equation}
    \tilde\Gamma_n=\tilde\Gamma_1\left[1+(n-1)^\delta\Delta\right]^\xi\,,
  \end{equation}
  \begin{equation}
        \tilde\rho_{n,\mathrm{prod}}=\tilde\rho_{1,\mathrm{prod}}\left[1+(n-1)^\delta\Delta\right]^\zeta\,.
  \end{equation}
\end{subequations}
We choose $\delta=1$ and $\Delta=0.1$ and perform a similar analysis
to the one presented above, for the cosmological parameters and
$(\tilde\Gamma_1,\tilde\rho_{1,\mathrm{prod}},\xi,\zeta)$. Here, the
relevance conditions introduced in \ref{sec:ambiguity} must also be
taken into account. They directly constrain the $\xi,\zeta$ parameter space for a given $N$. We choose $N=100$ as a compromise between
computing time to solve the system of equations and a large enough number of fields
that the decay times and initial densities are
distributed smoothly between the slowest and the fastest decaying
fields.	The results are encapsulated in Fig.~\ref{fig:tri-n100}. In
Fig.~\ref{fig:cdmvsn100} we show a comparison of the relevant
parameters with those obtained for the $\Lambda\mathrm{CDM}$ model. It is evident that this model does not address the Hubble tension in a more satisfactory way than the small $N$ models.

\begin{figure}\centering
\includegraphics[width=0.55\linewidth]{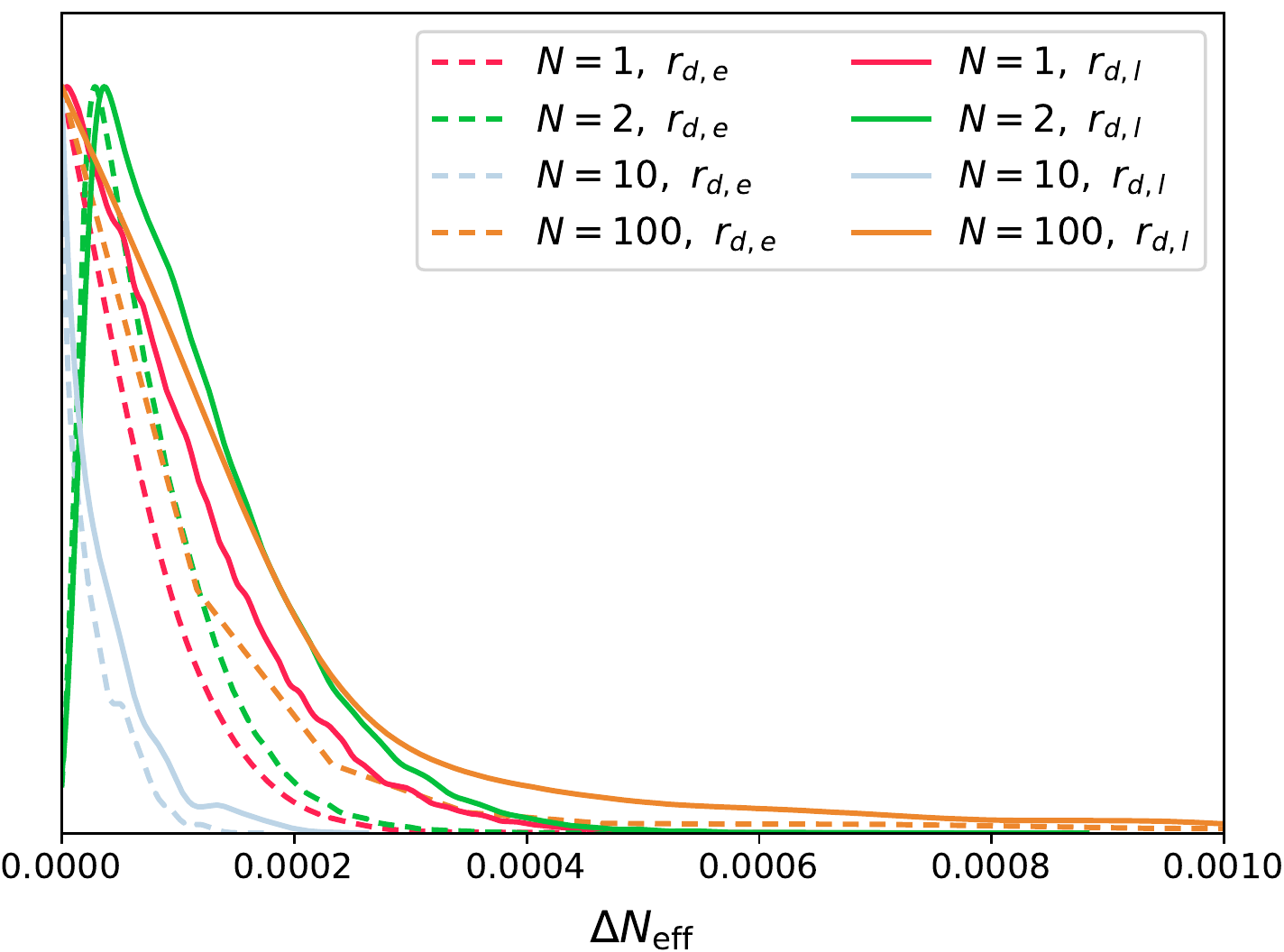}\caption{1D posterior distributions for $\Delta N_{\rm eff}$ for $N=1,2,10,100$.}\label{fig:neff}
\end{figure}

Finally, for completeness, in Fig.~\ref{fig:neff} we show the derived
posterior distribution for the calculated values of $\Delta N_{\rm
  eff}$. We see that in all the cases the contribution to $\Delta
N_{\rm eff}$ is comfortably below the bound given in
Eq.~(\ref{eq:deltaneff}). Note also that for $N=2$, the contribution to $\Delta
N_{\rm eff}$  is consistent with the value reported in model 1 of~\cite{Anchordoqui:2021gji}. 

\section{Conclusions}

To address the Hubble tension, we
examined dark sectors containing a large
number of decaying degrees of freedom with no trivial dynamics, with a focus on
decay processes that take place entirely among the dark
constituents. We further restricted ourselves to ensembles in which
CDM particles decay primarily to dark radiation in different
epochs. We showed that the data favor stable dark matter particles and
that a resolution of the $H_0$ tension with this type
of dark matter ensemble is elusive.

In closing, we comment on some interesting extensions that could
potentially evade our conclusion. Perhaps the most compelling of these
are models in which decays to final states that include
other relativistic massive particles occur. This allows for a dynamic equation of
state. It was recently shown in Ref.~\cite{Clark:2021hlo} that the
combination of such an $N=2$ model
and an early period of dark energy domination which reduces the linear
size of the sound horizon can ameliorate the $H_0$ tension to
within the 95\% CL. The early dark energy is modeled by a scalar field
that behaves like a cosmological constant at high redshifts ($z >
3000$) which then gets diluted at the same rate or faster than radiation
as the universe expands~\cite{Poulin:2018cxd}. We anticipate that
in principle, a similar reduction of the acoustic horizon may be obtained by enlarging the dark matter ensemble to allow
for very short-lived constituents that decay into particles that are
born relativistic but behave as CDM  before recombination. Our conclusion may also be evaded in models characterized by an ensemble in which the CDM
    particles decay into self-interacting dark radiation (as in {\it
      stepped fluids}~\cite{Aloni:2021eaq}), and models in which the ensemble couples to the dark energy sector through a quintessence field (as in string backgrounds with
  Standard Model fields confined on Neveu-Schwarz 5-branes~\cite{Anchordoqui:2019amx}).

\section*{Acknowledgements}
L.A.A. and J.F.S. are supported by the U.S.
National Science Foundation (NSF) under Grant No. PHY-2112527. V.B. is supported by the U.S. Department of Energy (DoE) under Grant No. DE-SC-0017647. D.M. is supported by the DoE under Grant No. DE-SC-0010504. D. M. thanks KITP, Santa Barbara for its hospitality and support via the NSF under Grant No. PHY-1748958 during the completion of this work.

\end{document}